\documentclass{emulateapj}
\usepackage{apjfonts}

\slugcomment{To appear in the ApJ Supp. Spitzer Issue}

\shorttitle{IRAC Camera for Spitzer}
\shortauthors{Fazio et al.}

\begin{document}

%% LaTeX will automatically break titles if they run longer than
%% one line. However, you may use \\ to force a line break if
%% you desire.

\title{The Infrared Array Camera (IRAC) for the Spitzer Space Telescope}

\author{G. G. Fazio\altaffilmark{1}, J. L. Hora\altaffilmark{1}, L. E.
Allen\altaffilmark{1}, M. L. N.  Ashby\altaffilmark{1}, P.
Barmby\altaffilmark{1}, L. K. Deutsch\altaffilmark{1,12}, J.-S.
Huang\altaffilmark{1}, S.  Kleiner\altaffilmark{1}, M.
Marengo\altaffilmark{1}, S. T. Megeath\altaffilmark{1}, G. J.
Melnick\altaffilmark{1}, M. A. Pahre\altaffilmark{1}, B. M.
Patten\altaffilmark{1}, J. Polizotti\altaffilmark{1}, H. A.
Smith\altaffilmark{1}, R. S. Taylor\altaffilmark{1}, Z.
Wang\altaffilmark{1}, S. P. Willner\altaffilmark{1}, W. F.
Hoffmann\altaffilmark{2}, J. L. Pipher\altaffilmark{3}, W. J.
Forrest\altaffilmark{3}, C. W.  McMurty\altaffilmark{3}, C. R.
McCreight\altaffilmark{4}, M. E. McKelvey\altaffilmark{4}, R. E.
McMurray\altaffilmark{4}, D. G. Koch\altaffilmark{4}, S. H.
Moseley\altaffilmark{5}, R. G. Arendt\altaffilmark{5}, J. E.
Mentzell\altaffilmark{5}, C. T. Marx\altaffilmark{5}, P.
Losch\altaffilmark{5}, P. Mayman\altaffilmark{5}, W.
Eichhorn\altaffilmark{5}, D. Krebs\altaffilmark{5}, M.
Jhabvala\altaffilmark{5}, D. Y. Gezari\altaffilmark{5},
D. J. Fixsen\altaffilmark{5}, J. Flores\altaffilmark{5}, K.
Shakoorzadeh\altaffilmark{5}, R. Jungo\altaffilmark{5}, C.
Hakun\altaffilmark{5}, L. Workman\altaffilmark{5}, G.
Karpati\altaffilmark{5}, R. Kichak\altaffilmark{5}, R.
Whitley\altaffilmark{5}, S. Mann\altaffilmark{5}, E.V.
Tollestrup\altaffilmark{6}, P. Eisenhardt\altaffilmark{7}, D.
Stern\altaffilmark{7}, V. Gorjian\altaffilmark{7}, B.
Bhattacharya\altaffilmark{8}, S. Carey\altaffilmark{8}, B. O.
Nelson\altaffilmark{8}, W. J. Glaccum\altaffilmark{8}, M.
Lacy\altaffilmark{8}, P. J. Lowrance\altaffilmark{8}, S.
Laine\altaffilmark{8}, W. T. Reach\altaffilmark{8}, J. A.
Stauffer\altaffilmark{8}, J. A. Surace\altaffilmark{8}, G.
Wilson\altaffilmark{8}, E. L.\altaffilmark{ }Wright\altaffilmark{9}, A.
Hoffman\altaffilmark{10}, G. Domingo\altaffilmark{10}, M.
Cohen\altaffilmark{11}}

\email{gfazio@cfa.harvard.edu}

\altaffiltext{1}{ Harvard-Smithsonian Center for Astrophysics, 60 Garden St.,
Cambridge, MA 02138}

\altaffiltext{2}{ Steward Observatory, Univ. of Arizona,
Tucson, AZ 85721}

\altaffiltext{3}{ Dept. of Physics and Astron., Univ. of Rochester,
Rochester, NY 14627}

\altaffiltext{4}{ NASA Ames Research Center, Moffett Field, CA 94035}

\altaffiltext{5}{ NASA Goddard Space Flight Center, Greenbelt, MD 20771}

\altaffiltext{6}{ Institute for Astronomy, 640 N. A'ohoku Pl., Hilo, HI 96720}

\altaffiltext{7}{ JPL, MS 264-767, 4800 Oak Grove Dr., Pasadena,
CA 91109}

\altaffiltext{8}{ Spitzer Science Center, Mail Code 220-6, California Institute of Technology,
Pasadena, CA 91125}

\altaffiltext{9}{ Dept. of Physics and Astronomy, Univ. of California at Los
Angeles, P.O. Box 951562, Los Angeles, CA 90095}

\altaffiltext{10}{Raytheon Infrared Operations, 75 Coromar Dr., Bldg. 2, MS 8, Goleta
, CA 93117}
\altaffiltext{11}{Univ. of California, 601 Campbell Hall, Berkeley, CA 94720}
\altaffiltext{12}{Deceased (4/02/04).}

%\newpage

\begin{abstract}
\label{sec:abstract}

The Infrared Array Camera (IRAC) is one of three focal plane instruments in
the Spitzer Space Telescope. IRAC is a four-channel camera that obtains
simultaneous broad-band images at 3.6, 4.5, 5.8, and 8.0 $\mu$m. Two nearly
adjacent 5.2$\times$5.2 arcmin fields of view in the focal plane are viewed
by the four channels in pairs (3.6 and 5.8 $\mu$m; 4.5 and 8 $\mu$m). All
four detector arrays in the camera are 256$\times$256 pixels in size, with
the two shorter wavelength channels using InSb and the two longer
wavelength channels using Si:As IBC detectors. IRAC is a powerful survey
instrument because of its high sensitivity, large field of view, and
four-color imaging. This paper summarizes the in-flight scientific,
technical, and operational performance of IRAC. 

\end{abstract}


\section{Introduction}

The three Spitzer Space Telescope focal plane instruments were designed 
to investigate
four major scientific topics: (1) early universe, (2) brown dwarfs and superplanets,
(3) active galactic nuclei, and (4) protoplanetary and planetary debris disks.
Of these topics, the most important in defining the Infrared Array
Camera (IRAC) design was 
the study of the 
early universe, and in particular, the study of the evolution of normal 
galaxies to z $>$ 3 by means of deep, large-area surveys. The 3 to 10 
$\mu$m wavelength range was selected because stars have a peak emission at 
a wavelength of 1.6 $\mu$m, at the minimum of the H$^-$ opacity \citep{john88}. 
The emission peak is an ubiquitous feature of stellar atmospheres and  can be 
used to determine a photometric redshift for 1 $ <$ z $<$ 5  \citep{wright94}.  
The IRAC sensitivity requirement was set such that IRAC
could achieve a 
10$\sigma$ detection of an $L^{*}$ galaxy at z = 3.  This 
in turn required the measurement of a flux 
density at 8 $\mu$m of 8 $\mu$Jy  \citep[10$\sigma$;][]{simpson99}.  
Channel 1 (3.6 $\mu$m) was selected to be at the minimum of the zodiacal 
background radiation \citep{wright85}, and to avoid the water ice absorption 
band at 3.1 $\mu$m.  The central wavelengths of the 
remaining IRAC filters and their bandwidths (approximately 25\%) were then 
optimized (considering the detector materials available) to reach the
sensitivity requirement and to permit the measurement 
of a photometric redshift from 1 $<$ z $<$ 5 \citep{simpson99}.

  The number of individual channels in IRAC was limited by the number of 
array cameras that could fit 
in the specified volume.  To reduce costs and maximize reliability, moving 
parts had to be minimized; hence no filter wheels, only fixed filters, were 
used.  Transmissive rather than reflective optics were used because 
of limited space.  
The pixel size was optimized to achieve the best point source sensitivity for
weak sources while maximizing the survey efficiency. 
The shutter at the entrance aperture was the only moving part allowed in IRAC.  

Although the design was optimized for the study of 
the early universe, IRAC is a general-purpose, wide-field camera that can be 
used for a large range of astronomical investigations.  In-flight observations 
with IRAC have already demonstrated that IRAC's sensitivity, pixel size, 
field of view, and filter selection are excellent for studying galaxy structure 
and morphology, active galactic nuclei, 
the early stages of star formation and evolution, and 
for identifying brown dwarfs.

IRAC was built by the NASA Goddard Space Flight Center (NASA/GSFC) with the
Smithsonian Astrophysical Observatory (SAO) having management and
scientific responsibility.
Additional information on the operation and performance of the IRAC
instrument can be found at the IRAC web site\footnote{
http://cfa-www.harvard.edu/irac}.

\section{Instrumentation Description}

IRAC consists of two parts: the Cryogenic Assembly (CA) installed in the
Multiple Instrument Chamber (MIC) within the Cryogenic Telescope Assembly
(CTA), and the Warm Electronics Assembly (WEA) mounted below the CTA in the
payload assembly area \citep{werner04}.

The IRAC Cryogenic Assembly, depicted in Figure 1, consists of the
following major subassemblies: two pickoff mirrors; the shutter; two optics
housings, which hold the doublet lenses, beamsplitters, filters, and cold
stops; four focal plane assemblies (FPAs) that include the detector arrays
and associated components; the transmission calibrator with its source and
integrating spheres; and the housing structure, consisting of the main
housing assembly and the wedge-shaped MIC adapter plate. The CA is cooled
to the temperature of the MIC base plate ($\sim $1.2 K). The CA has a mass
of 11.1 kg, is 0.15 m high and 0.28 m wide at its outer edge, and consumes an average 
of 3.0 mW of power during nominal operation. 

The IRAC WEA resides in Bay 5 of the spacecraft, operating at near room
temperatures ($\sim$10\degr C). All electrical interfaces between the
spacecraft and IRAC pass through the WEA. The WEA provides all power and
data interfaces to both the spacecraft and to the CA as necessary to
conduct the operation of IRAC. The WEA is connected to the CA via an
extensive set of conventional and cryogenic cables, which all pass through
a Junction Box located on the spacecraft near the CTA. The Junction Box
serves as the transition between the conventional and cryogenic cables.

The principal function of the WEA is the support and control of science
data taking: the generation of bias voltages to the detectors; the timing
of readout sequences; the amplification and digitization of the analog
science signals; the digital signal processing of the images; and the
transmission of the digital data to the spacecraft C{\&}DH solid state
recorders for mass storage, to be followed by downlinks to ground stations
at twelve hour intervals. Apart from autonomous fault protection, the WEA
internal software responds only to commands sent by the Spitzer Command
{\&} Data Handling (C{\&}DH)  computer. The WEA has a mass of 24.5 kg, and its 
dimensions are 0.46m in length, 0.38m in width, and 0.23m in height.  It
consumes 56 W of power during nominal operation.

\subsection{Optical Design}

The IRAC optical path is shown in Figure 2. Light enters the CA via two
pickoff mirrors located near the telescope focal plane. The two mirrors are
slightly displaced and tilted to physically separate the optical components
of the channel pairs. Thus the pickoff mirrors project 
nearly adjacent 5.2$\times$5.2 arcmin images of the sky to the channel
pairs. The centers of the two images are separated by 6.8 arcmin.

\begin{figure}
\includegraphics[clip,angle=-90,scale=0.39]{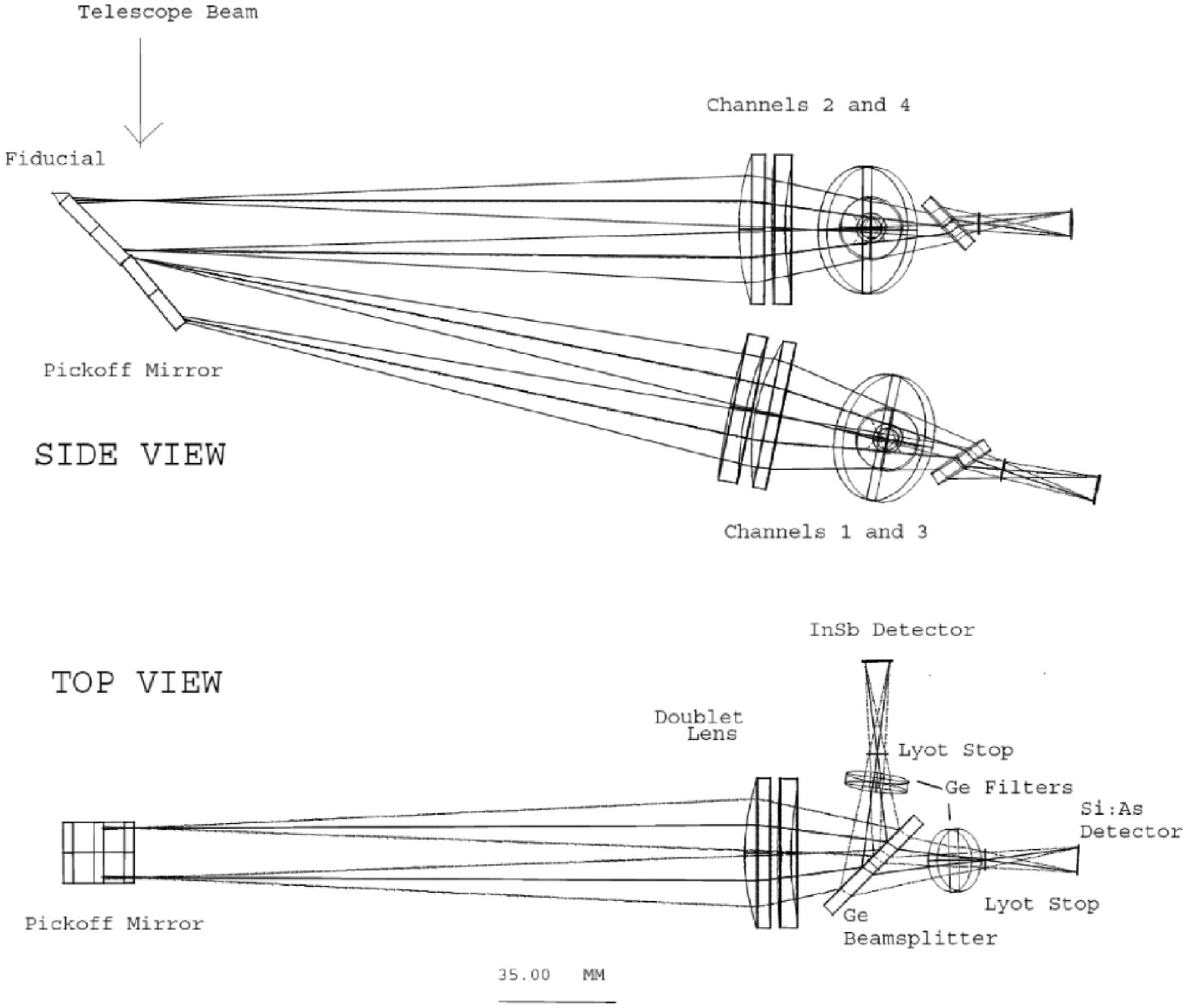}
\label{fig2}
\figurenum{2}
%\vspace{-1in}
\caption{ IRAC optical design, showing side view and top view.
}
\end{figure}

The lower mirror selects the field for channels 1 and 3. The reflected beam
is incident upon a MgF$_{2}$ -- ZnS (Cleartran) vacuum-spaced doublet that
re-images the Spitzer focal plane onto the detectors. After the lenses, a
germanium substrate beamsplitter reflects the channel 1 beam and transmits
the channel 3 beam. The beamsplitter is tilted by 45 degrees and is in the
converging beam. The channel 1 light is reflected from the beamsplitter,
passes through a bandpass filter, and then a Lyot stop, which rejects stray light. 
For channel 3,
after the beam goes through the beamsplitter, a filter 
selects the proper wavelength bandpass. The filter is
tilted at 45 degrees, in the plane opposite the beamsplitter, to correct
for the astigmatism introduced by the beamsplitter. A Lyot stop
rejects stray light.  The situation is similar for channels 2
and 4, which are selected by the upper pickoff mirror.  In these channels
the doublet lens materials are ZnSe and BaF$_{2}$. The channel 2 beam is
reflected by the beamsplitter, and the channel 4 beam is transmitted.  
All filters are mounted on germanium substrates. The Lyot stops are 
oversized by 15\%.  Barr Associates, Inc., Westford, MA, produced the filters,
OCLI-AJDS Uniphase Co., Santa Rosa, CA, the beamsplitters,  and Spectral
Systems, Inc., Dayton, Ohio, the lenses.

A shutter at the entrance to the CA was designed to block the incoming beams and to permit the measurement of
the dark count rate of the detector arrays.  A mirror, mounted on the rear surface of the shutter, 
reflects the light from the transmission calibration source into the two optical paths.
To eliminate the possibility that the shutter could get stuck in the closed
position, it has not been operated in flight to date.

IRAC has two internal calibration systems. The transmission calibration
system measures IRAC optical throughput, and the flood calibrators test for
detector responsivity and stability. Since the transmission calibration system
requires that the shutter be closed, it has not been used in flight.

\subsection{Focal Plane Arrays}

The IRAC detector arrays were developed by Raytheon/SBRC, Goleta, CA, under
contract to SAO \citep{hoff98, estr98}. Channels 1 (3.6 $\mu$m) and 2 (4.5
$\mu$m) InSb detector arrays operate at a temperature of 15 K, and Channels
3 (5.8 $\mu$m) and 4 (8.0 $\mu$m) Si:As detector arrays operate at 6 K. To help achieve
the relative photometric accuracy requirement of 2\%, the
array temperatures are controlled by an active feedback circuit to $<$ 10 mK
peak-to-peak variation from their set points.  Both array types are
256$\times$256 pixels in size, and have the same physical pixel size of 30
$\mu$m.  Cryo-CMOS technology was used in the readout multiplexer 
design \citep{wu97}.  
The arrays were
anti-reflection coated with SiO for Channels 1, 2, and 3, and with ZnS for
Channel 4.  The power dissipation for each array during nominal operations
is $<$ 1 mW. Table 1 lists some of the detector properties. The operability
is the percentage of the pixels in an array that meet specifications.

\begin{deluxetable}{llcccc}
\tablecaption{IRAC Detector Characteristics}
\tabletypesize{\scriptsize}
\tablewidth{0pt}
\tablehead{
& & & \colhead{Quantum} & \colhead{Well}\\
 &  &
\colhead{Read}  & \colhead{
Efficiency} & \colhead{Depth}  &
\colhead{Operability}\\
\colhead{Ch\#} & \colhead{FPA Designation} &
\colhead{Noise (e-)} & \colhead{(\%)} & \colhead{(e-)} &
\colhead{(\%)}
}
\startdata
1&
48534/34 (UR)&
8.1\tablenotemark{a}&
87&
145,000&
99.97\\
2&
48975/66 (GSFC)&
6.8\tablenotemark{a}&
86&
140,000&
99.99\\
3&
30052/41 (ARC)&
13.0\tablenotemark{a}&
45&
170,000&
99.99\\
4&
30219/64 (ARC)&
6.6\tablenotemark{b}&
70&
200,000&
99.75\\
\enddata
\tablenotetext{a}{200 s frame time}
\tablenotetext{b}{50 s frame time}
\label{tab1}
\end{deluxetable}

The arrays can also be heated above their nominal operating temperatures to
perform a thermal ``anneal" of the FPAs to remove hot pixels or residual
images.  In the in-flight anneal operation, the temperatures of the FPAs are
raised to 23 K (Channel 1) and 30 K (Channels 2,3,4) for approximately 2
minutes and then returned to their nominal operating temperatures in 
approximately 4 minutes.

The InSb arrays were tested and characterized at the University of
Rochester \citep{pipher00, benson00} and the Si:As arrays at the NASA/ARC
\citep{mcmurray00}. The flight arrays were evaluated at the NASA/GSFC.

\subsection{IRAC Operation Mode}

IRAC has one method of operation on the Spitzer Space Telescope: stare and
integrate. The field-of-view is determined by use of the full array
(256$\times$256 pixels) or the sub-array (32$\times$32 pixels) readout 
mode. Frame times
for the full array can vary from 0.4 to 500 sec in 0.2 sec steps, but the following times
are standard: 2, 12, 30, 100, and 200 sec, except that channel 4 uses two
or four 50-sec frames instead of 100 and 200 sec frames, respectively.

In the  subarray mode, a 
32$\times$32 pixel region is read out at a faster rate. The standard frame
times are: 0.02, 0.1, and 0.4 sec; other times can be commanded in steps
of 10 msec.  Each data collection command in subarray mode produces 64 image 
frames, taken consecutively with no dead time between frames. 
This mode is primarily for observing
bright sources that would otherwise saturate the array, or for observing
time-critical events where a finer time resolution is required. This mode
was used to measure the pointing jitter of the telescope.

During an integration data are taken using a Fowler Sampling method  \citep{fowler90}  to
reduce the effective read noise. This mode of sampling consists of taking N
non-destructive reads immediately after the reset (pedestal levels) 
and another N
non-destructive reads near the end of the integration (signal levels).  
The average 
of the pedestal levels is subtracted from the average of the signal levels to derive 
the signal. The averaging and subtraction of the two sets of reads is
performed in the IRAC on-board electronics to generate a single image from
each array, which is stored and transmitted to the ground. N can vary from
1 to 64. The Fowler N used for an observation will depend on integration
time and has been selected to maximize the S/N at each frame time, based on
in-flight performance tests.

IRAC is often used in a high dynamic range mode where a long exposure is preceded
by one or more short exposures: 0.6, 12 sec; 1.2, 30 sec; 0.6, 12, 100 sec;
and 0.6, 12, 200 sec.
IRAC is capable of operating each of its four arrays independently and/or
simultaneously. In normal flight operation, all four arrays are operated
together.

\section{Sensitivity}

The point source sensitivities for the frame times available to IRAC 
are shown in Table 2.  For each frame time and channel, the 1$\sigma$
sensitivity in $\mu$Jy is given for the low background case (near the ecliptic
poles).  The values were calculated based on the sensitivity model given by
\citet{hora00}, using measured in-flight values for the read noise,
pixel size, noise pixels, background, and total system throughput.  The
numbers were compared to observations at several frame times to confirm the
validity of the calculations.  The values in Table 2 are close to
pre-flight predictions of 0.5, 0.9, 3.1, and 5.0 $\mu$Jy (1$\sigma$, 200 sec) 
for channels 1-4, respectively, 
except for the influence of two factors: the
lower (better) than expected noise pixel values for all channels, and the
lower (worse) throughput in channels 3 and 4.  IRAC meets its
required sensitivity limits of 0.92, 1.22, 6, and 9 $\mu$Jy (1$\sigma$, 200 sec) for
channels 1-4, respectively.

\begin{deluxetable*}{crrrr}
\tablecaption{IRAC Point Source Sensitivity (1$\sigma$, $ \mu $Jy, low
background)}
\tabletypesize{\scriptsize}
\tablewidth{0pt}
\tablehead{
\colhead{Frame Time} & \\
\colhead{(sec)} & \colhead{3.6 $\mu $m} &
\colhead{4.5 $\mu $m}  & \colhead{5.8 $\mu $m} & \colhead{8.0 $\mu $m}
}
\startdata
200&
0.40&
0.84&
5.5&
6.9 \\
100&
0.60&
1.2&
8.0&
9.8 \\
30&
1.4&
2.4&
16&
18 \\
12&
3.3&
4.8&
27&
29 \\
2&
32&
38&
150&
92 \\
0.6\tablenotemark{a}&
180&
210&
630&
250 \\
0.4\tablenotemark{b}&
86&
75&
270&
140 \\
0.1\tablenotemark{b}&
510&
470&
910&
420 \\
0.02\tablenotemark{b}&
7700&
7200&
11000&
4900\\
\enddata
\tablenotetext{a}{high dynamic-range mode}
\tablenotetext{b}{subarray mode}
\label{tab2}
\end{deluxetable*}

The noise measurements in channels 3 and 4 scale as expected 
with the inverse square root of time to the limit measured, 
20,000 seconds.   Channels 1 and 2 
deviate from this rule in approximately 3000 seconds, which indicates that 
these channels are approaching the source completeness confusion limit
\citep[see also][]{fazionum04}. 
The deviation occurs at about 1 $\mu$Jy (5$\sigma$), which is slightly fainter
than the predictions for the IRAC confusion limit \citep{vaisanen01}. 

For extended emission in channels 3 and 4, the calibration was found to be 
significantly different than
expected from the calibration derived from a point source. The calibration
aperture (10 pixel radius;  12\farcs2 aperture radius) does not capture all
of the light from the calibration sources so the extended sky emission
appears too bright. For photometry using different apertures, the estimated
correction is listed in Table 6.4 of the Spitzer Observer's
Manual\footnote{http://ssc.spitzer.caltech.edu/documents/som/}. 

\section{IRAC Image Quality}

IRAC's optics provide diffraction-limited imaging, with wavefront errors $<
\lambda / 20$ in each channel. The in-flight IRAC image quality is limited by a 
combination of the optical quality of the Spitzer telescope, which is diffraction 
limited at 5.4 $\mu$m, and  the size of the IRAC pixels.  The predicted FWHM of the 
point spread function (PSF), based on the preflight telescope and IRAC optical models, was 1.6, 1.6, 1.8,  and 1.9 arcsec at 3.6, 4.5, 5.8, and 8.0 $\mu$m, respectively.  Table 3 shows the properties of the
IRAC PSF, derived from in-flight measurements of
bright stars.   The FWHM of the PSF will differ slightly for non-stellar
spectra, e.g. dust.
The PSF was monitored throughout the early mission while the telescope cooled down, and while the secondary mirror was moved in two
steps to achieve a near-optimal focus \citep{hoff03}. The ``noise pixels''
column in Table 2 gives the equivalent number of pixels whose noise
contributes to noise in the analysis when an image is spatially filtered
for optimum faint point-source detection \citep{king83}.

\begin{deluxetable*}{ccccccc}
\tablecaption{IRAC Image Quality}
\tabletypesize{\scriptsize}
\tablewidth{0pt}
\tablehead{
& \colhead{Noise} & \colhead{FWHM} & \colhead{FWHM of} & \colhead{Central} &
\colhead{Mean Pixel} & \colhead{Maximum} \\
& \colhead{Pixels} &  \colhead{(mean; } & \colhead{centered} & \colhead{pixel fl
ux} &
\colhead{Scale} & \colhead{Distortion} \\
\colhead{Chan.} & \colhead{(mean)} & \colhead{arcsec)} & \colhead{(arcsec)} &
\colhead{(peak; \%)} & \colhead{(arcsec)} &
\colhead{(pixels)\tablenotemark{a}}
}
\startdata
1&
7.0&
1.66&
1.44&
42&
1.221&
1.3\\
2&
7.2&
1.72&
1.43&
43&
1.213&
1.6\\
3&
10.8&
1.88&
1.49&
29&
1.221&
1.4\\
4&
13.4&
1.98&
1.71&
22&
1.220&
2.2\\
\enddata
\tablenotetext{a}{Maximum distortion in pixels relative to a square grid}

\end{deluxetable*}

There are two columns for the full width at half-maximum (FWHM) of the PSF.
The mean FWHM is from observations of a star at 25 different location on
the array. The FWHM for ``centered PSF'' is for cases where the star was
most closely centered in a pixel. The fifth column in Table 2 is the
percentage of the total flux in the central pixel for a point source that is well-centered
in a pixel. The flux in the central pixel for a random observation will be
lower, because the PSF of the telescope is rather undersampled at the IRAC
pixel scale except in channel 4.  The Strehl Ratios observed are 0.37, 0.58, 0.62, and 0.95 for channels 1-4 respectively.  The Strehl Ratio for channel 3, after correction for scattered light in the array detector, is 0.71.

\subsection{Distortion}

There is a
small amount of distortion over the IRAC FOV (Table 3).  A quadratic
distortion model is provided in the Spitzer Science Center (SSC) Basic
Calibrated Data (BCD) pipeline which fits in-flight data at
rms $<$0\farcs2 (channels 1 and 2) and $<$0\farcs3 (channels 3 and 4)
across the full arrays.

\subsection{Scattered and Stray Light}

Diffuse stray light from outside the IRAC FOV is scattered into the active
region of the IRAC detectors in all four channels. The problem is more
significant in channels 1 and 2 than in channels 3 and 4.  In channels 1
and 2 the stray light pattern due to diffuse sources resembles a
``butterfly'', while in channels 3 and 4 it resembles a ``tic-tac-toe''
board.  A discussion of these effects and example images are given by
\citet{hora04}.  
These artifacts are due to zodiacal light scattered onto the arrays,
possibly reflected from a hole in the FPA covers in channels 1 and 2, and
from reflective surfaces outside the edges of channel 3 and 4 arrays.  The
stray light scales with zodiacal light, which is the light source for the
flat fields, so the stray pattern contaminates the flats and all IRAC
images. The "butterfly wings" have amplitudes of about 5{\%} of the background intensity.
The pipeline subtracts a model for the tic-tac-toe and butterfly
patterns for the data, scaled based on a model for the zodiacal light
intensity \citep{kelsall98} at the location and time of the observation.  
To the extent the model is accurate, this corrects for the stray light
contamination.

Stars that fall in specific regions just outside the array edges scatter light into the detectors and produce
distinctive patterns of scattered light on the array. Scattered light
avoidance zones have been identified in
each channel. Observers should avoid placing bright stars in these zones if their
observations are sensitive to scattered light. Typically, in channels 1 and
2 about 2{\%} of the light from a star is scattered into a ``splatter
pattern'', which has a peak intensity of about 0.2{\%} of the light from
the star. The avoidance zones for channels 3 and 4 are a narrow strip about
3 pixels wide, 16 pixels outside of the array and surrounding it.

\section{Calibration}

The overall requirement for the IRAC mission is that the system photometric
responsivity be calibrated to a relative accuracy of 2{\%} and that the
absolute accuracy of the data set be determined to better than 10{\%}. We
expect that both of these requirements will be achieved. 
Currently 25{\%} of the time that IRAC is on is devoted to
calibration, but the excellent stability of IRAC will result in a
significant decrease in that fraction. 

\subsection{System Throughput and Isophotal Wavelength}

The IRAC system throughput and optical performance are governed by a
combination of the system components, including the lenses, beamsplitters,
filters, mirrors, and detectors. Figure 3 gives the system
throughput, including transmission 
of the telescope, IRAC optics, and quantum efficiency of
the detectors \citep{hora01, stewart00}. For each IRAC channel, both
polarizations were measured \citep{stewart00}, and the average polarization
is shown as a solid curve.

\begin{figure}
\includegraphics[angle=-90,scale=0.33]{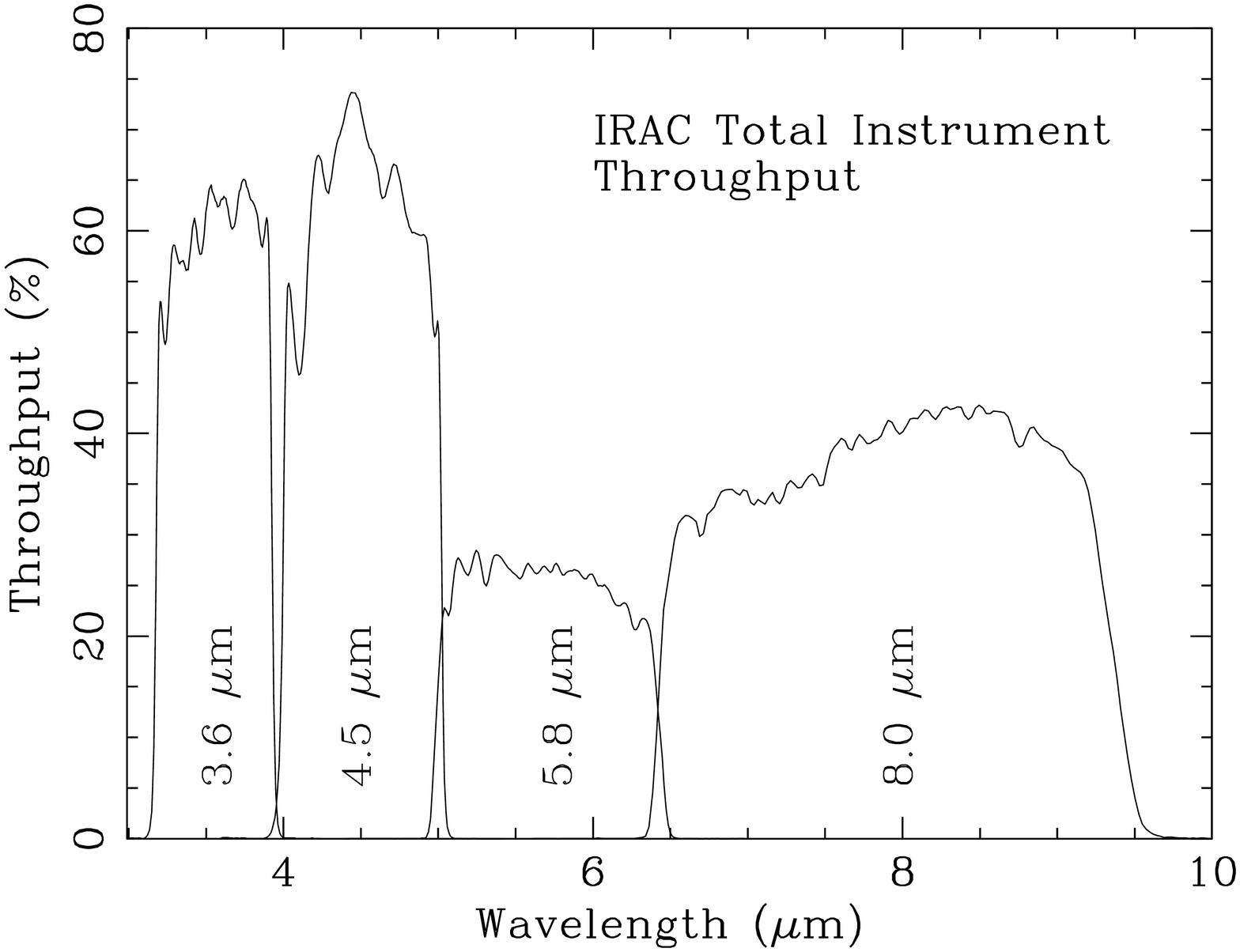}
\label{fig3}
\figurenum{3}
\caption{
IRAC system total throughput, including transmission of the telescope, IRAC
optics, and quantum efficiency of the detectors.
}
\end{figure}

The system parameters are summarized in Table 4. The isophotal wavelength
$E(\lambda_{isophotal})$ is the wavelength which must be assigned to the monochromatic flux density
derived from a broadband measurement.  The isophotal wavelength is 
defined as 

\begin{equation}
E = \int_{a}^{b}E(\lambda) T(\lambda) d\lambda = E(\lambda_{isophotal})
\int_{a}^{b} T(\lambda) d\lambda
\end{equation}

where $E$ is the measured in-band flux, $E(\lambda)$ is the irradiance at the 
entrance aperture of the telescope, $T(\lambda)$ is the spectral
response of the system, and $a$ and $b$ define the wavelength range of interest.
 We calibrate the flux
densities for a nominal spectrum $\nu I_{\nu }$ = constant. For source
spectra with a range of slopes, $\alpha = d\log F_{\nu }/d\nu $ from --2 to
+2, the color corrections are very close to unity (0.99 to 1.02). A color
correction table for various values of the spectral index will be given in
the Spitzer Observer's Manual.

\begin{deluxetable*}{cccccccc}
\tablecaption{IRAC Channel Characteristics}
%\tabletypesize{\scriptsize}
\tablehead{
&\colhead{Isophotal $\lambda$\tablenotemark{a}}&
\colhead{Center $\lambda$\tablenotemark{b}} &
\multicolumn{2}{c}{Bandwidth\tablenotemark{c}}&
\colhead{Average}&
\colhead{Min. In-band}&
\colhead{Peak} \\
\colhead{Channel} & \colhead{($\mu $m)} & \colhead{($\mu $m)} &
\colhead{($\mu $m)} &
\colhead{(\%)} & \colhead{Transmission} & \colhead{Transmission} & \colhead{Tran
smission}
}

\startdata
1&
3.535&
3.58&
0.750&
21&
0.676&
0.563&
0.748 \\
2&
4.502&
4.52&
1.015&
23&
0.731&
0.540&
0.859 \\
3&
5.650&
5.72&
1.425&
25&
0.589&
0.552&
0.653 \\
4&
7.735&
7.90&
2.905&
36&
0.556&
0.450&
0.637\\
\enddata
\tablenotetext{a}{The isophotal wavelength; see definition in text. }
\tablenotetext{b}{The center wavelength is the midpoint between the points on
the transmission curve that define the bandwidth.}
\tablenotetext{c}{The bandwidth is the full width of the band at 50\% of the
average in-band transmission.}
\end{deluxetable*}

\subsection{Absolute Calibration}

A number of astronomical standard stars are observed in each instrument
campaign to obtain a relative flux calibration \citep{megeath01}.
Stars with a range of spectral indices and fluxes are observed at a number
of positions across the array and many times throughout the mission to
monitor any changes that may occur. To calculate Vega magnitudes from the
calibrated IRAC images, use the following zero-magnitude fluxes: 277.5,
179.5, 116.6, and 63.1 Jy for channels 1, 2, 3, and 4, respectively.  These fluxes were calculated
from the IRAC throughput values and the absolute spectra at IRAC wavelengths  \citep{cohen03}.

\subsection{Stability}

In-flight data indicate that IRAC is very stable; observations of
calibration stars taken over two months are repeatable to about 2{\%}.

\subsection{Linearity}

Both types of detectors have measurable non-linearity.  To linearize the
data, a quadratic fit is used for the InSb arrays, and a cubic fit is used
for the Si:As arrays. All arrays have been linearized to better than 1{\%}
up to approximately 90{\%} of their full-well capacity. The detector
linearity was measured during ground testing and verified during flight.

\subsection{Flat Field}

The flat field for each channel is defined as the factors by which one must
correct each pixel to give the same value at a particular flux for uniform
illumination. The factors include any field-dependent optical transmission
effects as well as the responsivity of the pixel and output electronics
gain.  Sky flats are obtained using a network of 24 high zodiacal
background regions of the sky in the ecliptic plane,  ensuring a
relatively uniform illumination with a reasonable amount of flux.  If the
data are taken with an appropriate dither pattern, it is possible to relate
the total response of each pixel to that of all others. One such region is
observed at the beginning and end of each  IRAC campaign. The data are reduced
and combined to construct a 
flat field image for each channel, which is then divided into the science data.

\subsection{Dark Frames, Offsets and ``First Frame'' Effect}

The detector dark currents are generally insignificant compared to the sky
background. However, there is a significant offset or bias (which can be
positive or negative) in a dark frame, which must be
subtracted from the observations. Especially for shorter frames, the
``dark'' images are mostly due to electronic bias differences, rather than
true dark current. Therefore, the number of electrons in a dark image does
not scale linearly with exposure time. The shutter has never been used in
flight, and therefore isolated dark/bias data cannot be taken.  Instead,
ground-based ``lab'' darks are used, in conjunction with sky darks which 
measure changes to the darks that have occurred since launch.

As part of routine operations a dark region of the sky near the north
ecliptic pole is observed at the beginning and end of an IRAC campaign, and  
every three days during the campaign. 
These data are reduced and combined in such a way as to reject stars and
other astronomical objects with size scales smaller than the IRAC array.
The resulting image of the minimal uniform sky background contains both the
bias and the dark current. When subtracted from the routine science data,
this will eliminate both of these instrumental signatures.  The dark frames, when corrected for
changes in the zodiacal background flux, remain very stable from campaign to campaign.

In general, the first frame of any sequence of images tends to have a
different median value from other frames in the sequence, hence the name ``first frame effect.'' 
The largest offsets occur when a frame is taken with a very short
interval from the preceding image, which occurs when multiple frames are
commanded at once. Such commands have been eliminated from IRAC science 
observations, so that there is always a finite interval between one frame and its predecessor.
Since many IRAC observations will take only one image
at each sky position, and there is an 8 to 60 second delay while Spitzer
points and settles to a new position, the "first frame effect" is reduced significantly. 
The correction for the first frame effect was derived from pre-flight ground-based
 data.

\subsection{Cosmic Ray Effects}

In flight, cosmic rays excite approximately 3 pixels per second in channels 1 and 2
and approximately 5 pixels per second in channels 3 and 4. This is
close to the predicted rate \citep{mason83}. The most common cosmic rays do not affect
the pixel performance in subsequent frames and are confined to a few pixels
around the peak. The cosmic ray effects in channel 1 and 2 are more compact than
those in channels 3 and 4, which have thicker detectors. In these latter
channels the cosmic ray effects appear as streaks and blobs. However, other
events can cause streaks or other multiple pixel structures in the array,
and less common energetic and high-Z events can cause image artifacts in
subsequent frames. Most of the cosmic ray effects are removed by the
post-BCD pipeline processing for observations that are well dithered.

On 2003 October 28, Spitzer encountered a large solar proton flare with an
integrated dose of 1.6 x 10$^{9}$ p/cm$^{2}$ for proton energies greater
than 50 MeV. IRAC was powered off at the time. After several thermal
anneals of the arrays following the flare, only a few ($\sim$20)  new hot
pixels were detected, all in the channel 4 array.

\section{Artifacts of the IRAC Arrays}

Array detector artifacts known before launch or discovered in flight
include: persistent images in channels 1 and 4, multiplexer bleed, column
pulldown, and banding. These effects are described in more detail in the
SSC Observer's Manual and in articles by \citet{hora00, hora01, hora03,
hora04}.
To erase persistent images both channel 1 and 4 are annealed simultaneously
every 12 hours (after each downlink) and bright source observations are
scheduled only near the end of an IRAC campaign. Column pulldown, bright
source multiplexer bleed, and banding will be removed in the post-BCD
pipeline corrections.  The effects of persistent images and cosmic rays can
be avoided or greatly reduced in observations that are well dithered. We
recommend taking a minimum of three images of a science target in each
channel as a best practice for observing with IRAC.

\section{Summary}

In flight, IRAC meets all of the science requirements which were established before launch, and in many cases the performance is better than the requirement.  IRAC is a powerful survey instrument because of its high sensitivity, large field of view, and four-color imaging from 3.2 to 9.5 microns wavelength.  
For example, in channel 1 (3.6 $\mu$m), IRAC can reach the same point source
sensitivity (19 magnitude, 5$\sigma$, 1 hour) as the W. M. Keck telescope at 2.9 - 3.2 $\mu$m
in only one-hundredth the exposure time and does so with a far larger
field of view: 5\arcmin$\times$5\arcmin~ versus 40\arcsec$\times$40\arcsec.
In channel 4 (8.0 $\mu$m), IRAC can achieve the ultimate sensitivity 
limit reached by ISOCAM (15 $\mu$Jy, 1$\sigma$, 1 hour, 3\arcmin$\times$3\arcmin~ field) on the Infrared 
Space Observatory (ISO) in the LW2 filter (6.7 $\mu$m) in approximately 
one-hundreth the frame time \citep{alteri00}.  

IRAC continues to function extremely well since the
first images were produced, seven days after launch, on 2003 September 1.  The first scientific results of IRAC, which are presented in this issue, are examples of IRAC's great potential for producing exciting new science with the Spitzer Space Telescope.

\begin{acknowledgements}
Lynne Deutsch, a co-author of this paper, died on 2 April 2004 after a long 
illness.  Lynne was a dear friend and a close colleague.  The IRAC team will
deeply miss her presence. We dedicate this paper in her memory for all her 
contributions to infrared astronomy.

This work is based on observations made with the Spitzer Space Telescope,
which is operated by the Jet Propulsion Laboratory, California Institute of
Technology under NASA contract 1407. Support for this work was provided by
NASA through Contract Number 125790 issued by JPL/Caltech.

Support for the IRAC instrument was provided by NASA through Contract
Number 960541 issued by JPL.

\end{acknowledgements}

\clearpage

\begin{figure}
\plotone{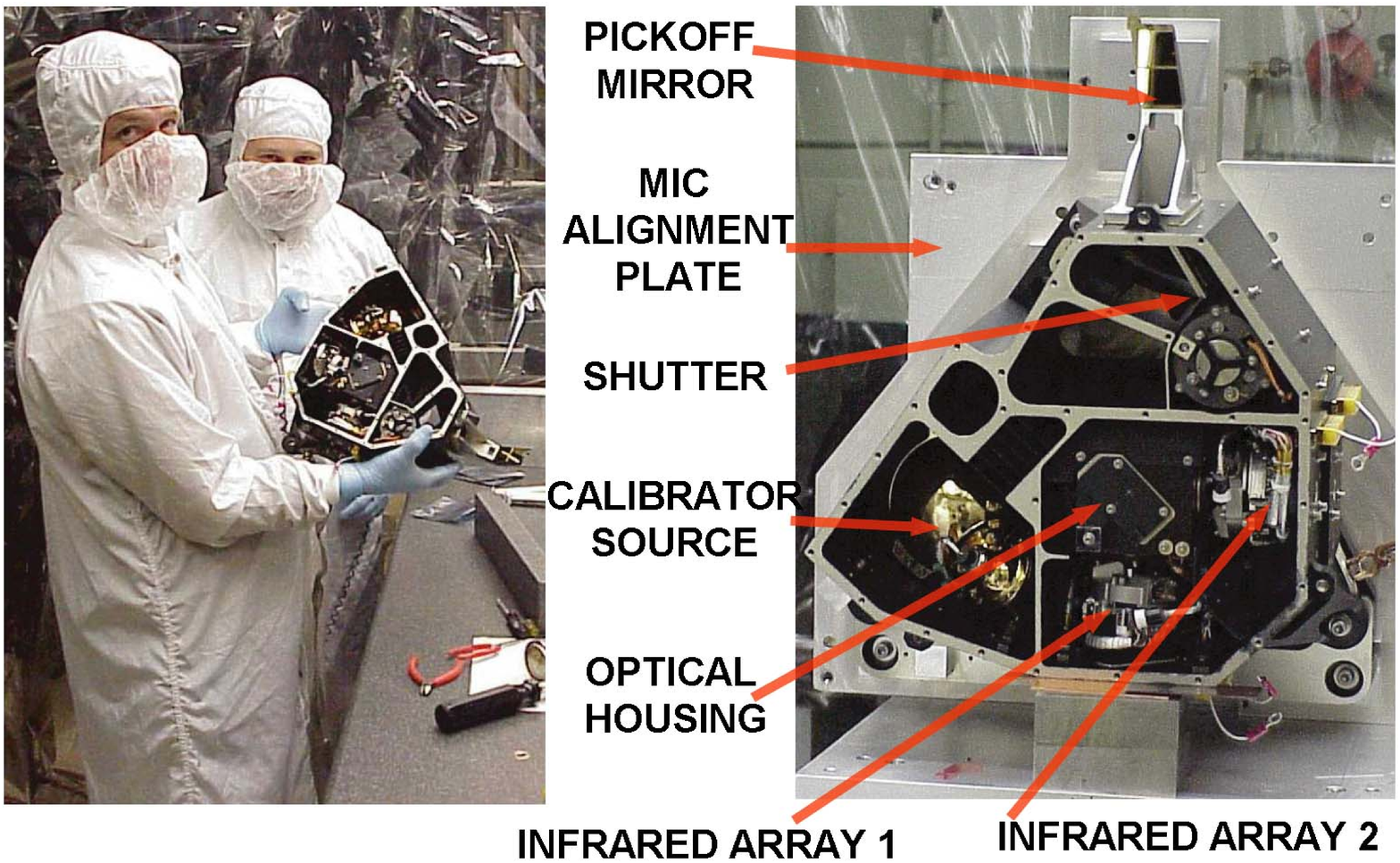}
\figurenum{1}
\label{fig1}
\caption{
IRAC Cryogenic Assembly at NASA Goddard Space Flight Center,
with the top cover removed to show the inner components. The MIC alignment plate
was used only for testing.  The parts marked Infrared Array 1 and 2 are the IRAC
Channel 4 and 2 focal plane assemblies, respectively.
}
\end{figure}


\begin{thebibliography}{}

\bibitem[Alteri et al.(2000)]{alteri00} Alteri, B. Metcalfe, L., Blommaert, J., Elbaz, D. Starck, J.-L., {\&} Aussel, H. 2000, Exp. Astr., 10, 291

\bibitem[Benson et al.(2000)]{benson00} Benson, R. G., Forrest, W. J.,
Pipher, J. L., Glaccum, W., Solomon, S. L. 2000, in \textit{Infrared
Spaceborne Remote Sensing VIII}, ed s. M. Stojnik {\&} B. F. Andresen,
Proc. SPIE, 4131, 171

\bibitem[Cohen et al.(2003)]{cohen03} Cohen, M., Megeath, S. T., Hammersley, P. L., {\&} Stauffer, J. 2003, \aj, 125, 2645

\bibitem[Estrada et al.(1998)]{estr98} Estrada, A. D. et al. 1998, in
\textit{Infrared Astronomical Instrumentation}, ed. A. Fowler, Proc. SPIE
3354, 99

\bibitem[Fazio et al.(2004)]{fazionum04}  Fazio, G. G.,  Ashby, M. L. N., Barmby,
P., Hora, J. L.,  Huang, J.-S.,  Pahre, M. A.,  Wang, Z.,
Willner, S. P., Arendt, R. G., Moseley, S. H., Brodwin, M.,  Eisenhardt, P.,
Stern, D., 
Tollestrup, E. V., \$  Wright, E. L. 2004, \apjs, in press 

\bibitem[Fowler \& Gatley(1990)]{fowler90} Fowler, A. M. and Gatley, I. 1990, \apjlett, 353, 33

\bibitem[Hoffman et al.(1998)]{hoff98} Hoffman, A. W., et al. 1998, in
\textit{Infrared Astronomical Instrumentation}, ed. A. F owler, Proc. SPIE,
3354, 24

\bibitem[Hoffmann et al.(2003)]{hoff03} Hoffmann, W. F., Hora, J. L.,
Mentzell, J. E., Trout-Marx, C., Eisenhardt, P. R. 2003 , in
\textit{Infrared Space Telescopes and Instruments}, SPIE Proc., 4850, 428

\bibitem[Hora et al.(2000)]{hora00} Hora, J. L., et al. 2000, in
\textit{Infrared Spaceborne Remote Sensing VIII}, eds. M. Stojnik {\&} B.
F. Andresen, Proc. SPIE, 4131, 13

\bibitem[Hora et al.(2001)]{hora01} Hora, J. L., et al.
 2001, in ``\textit{The Calibration Legacy of the ISO Mission}'', VILSPA,
Spain, February 2001, (ESA SP-481), 73 

\bibitem[Hora et al.(2003)]{hora03} Hora, J. L., et al. 2003, in
\textit{Infrared Space Telescopes and Instruments}, SPIE Proc., 4850, 83

\bibitem[Hora et al.(2004)]{hora04} Hora, J. L., et al. 2004, in
\textit{Optical, Infrared, and Millimeter Space Telescopes}, SPIE Proc. 5487,
in press

\bibitem[John(1988)]{john88} John, T. L. 1988, \aap, 193, 189

\bibitem[Kelsall et al.(1998)]{kelsall98}Kelsall, T. et al., \apj, 508, 44

\bibitem[King(1983)]{king83}King, I. R. 1983, \pasp, 95, 163

\bibitem[Mason \& Culhane(1983)]{mason83} Mason, I. M. and Culhane, J. L., IEEE Transactions in 
Nuclear Science, NS-30, No. 1, 485

\bibitem[Megeath et al.(2001)]{megeath01} Megeath, S. T., Cohen, M.,
Stauffer, J., Hora, J. L., Fazio, G., Berlind, P., {\&} Calkins, M. 2001,
in ``\textit{The Calibration Legacy of the ISO Mission}'', VILSPA, Spain,
February 2001, (ESA SP-481), 165

\bibitem[McMurray et al.(2000)]{mcmurray00} McMurray, R. E., et al. 2000,
in \textit{Infrared Spaceborne Remote Sensing VIII, }eds. M. Stojnik {\&}
B. F. Andresen, Proc. SPIE, 4131, 62

\bibitem[Pipher et al.(2000)]{pipher00} Pipher, J. L., et al. 2000, in
\textit{Infrared Spaceborne Remote Sensing VIII}, eds. M. Stojnik {\&} B.
F. Andresen, Proc.SPIE, 4131, 7

\bibitem[Simpson \& Eisenhardt(1999)]{simpson99} Simpson, C. and Eisenhardt, P. 1999, \pasp, 111, 691

\bibitem[Stewart \& Quijada(2000)]{stewart00} Stewart, K. P., and Quijada,
M. A. 2000, in \textit{Infrared Spaceborne Remote Sensing VIII, }eds. M.
Stojnik {\&} B. F.  Andresen, Proc. SPIE, 4131, 2000

\bibitem[V\"ais\"anen et al.(2001)]{vaisanen01} V\"ais\"anen, P., Tollestrup, E. V., {\&} Fazio, G. G. 2001, \mnras,
325, 1241

\bibitem[Werner et al.(2004)]{werner04} Werner, M. et al. 2004, \apjsupp,
this issue

\bibitem[Wright(1985)]{wright85} Wright, E. L. 1985, \pasp, 97, 451.

\bibitem[Wright et al.(1994)]{wright94} Wright, E. L., Eisenhardt, P., and Fazio, G. 1994, \baas, 184, 2503

\bibitem[Wu et al.(1997)]{wu97} Wu, J., Forrest, W. J., Pipher, J. L., Lum, N., {\&} Hoffman, A. 1997, Rev. of Sci. Inst., 68, 3566

\end{thebibliography}
\end{document}